\documentclass[fleqn,usenatbib]{mn2e}

\usepackage[a4paper]{geometry}
\usepackage[T1]{fontenc}
\usepackage{ae,aecompl}

\usepackage{natbib}
\usepackage{graphicx}	% Including figure files
\usepackage{amsmath}	% Advanced maths commands
\usepackage{amssymb}	% Extra maths symbols

\usepackage{color}
\definecolor{red}{rgb}{0.7,0.1,0.1}

\definecolor{blue}{rgb}{0.1,0.1,0.7}

\usepackage{hyperref}

\usepackage{subfigure}

\title[Baryonic Effects on Planes of Satellites]{The Role of Baryons in Creating Statistically Significant Planes of Satellites around Milky Way-Mass Galaxies}

\author[Ahmed et al.]{
Sheehan H. Ahmed,$^1$\thanks{E-mail: \href{mailto:shahmed@physics.rutgers.edu}{shahmed@physics.rutgers.edu}}
 Alyson M. Brooks,$^1$\thanks{E-mail: \href{mailto:abrooks@physics.rutgers.edu}{abrooks@physics.rutgers.edu}}
 Charlotte R. Christensen$^2$\thanks{E-mail: \href{christenc@grinnell.edu}{christenc@grinnell.edu}}
\\
\vspace*{6pt} \\
$^1$Department of Physics \& Astronomy, Rutgers, The State University of New Jersey, 136 Frelinghuysen Rd, Piscataway, NJ 08854\\
$^2$Department of Physics, Grinnell College, Noyce Science Center, 1116 Eighth Ave, Grinnell, IA 50112 
}

\date{10 Oct, 2016}
\pubyear{2016}

\begin{document}
\label{firstpage}
\pagerange{\pageref{firstpage}--\pageref{lastpage}}
\maketitle

\begin{abstract}

We investigate whether the inclusion of baryonic physics influences the formation of thin, coherently rotating planes of satellites such as those seen around the Milky Way and Andromeda.  For four Milky Way-mass simulations, each run both as dark matter-only and with baryons included, we are able to identify a planar configuration that significantly maximizes the number of plane satellite members.  The maximum plane member satellites are consistently different between the dark matter-only and baryonic versions of the same run due to the fact that satellites are both more likely to be destroyed and to infall later in the baryonic runs. Hence, studying satellite planes in dark matter-only simulations is misleading, because they will be composed of different satellite members than those that would exist if baryons were included. Additionally, the destruction of satellites in the baryonic runs leads to less radially concentrated satellite distributions, a result that is critical to making planes that are statistically significant compared to a random distribution. Since all planes pass through the centre of the galaxy, it is much harder to create a plane from a random distribution if the satellites have a low radial concentration.  We identify Andromeda's low radial satellite concentration as a key reason why the plane in Andromeda is highly significant. Despite this, when co-rotation is considered, none of the satellite planes identified for the simulated galaxies are as statistically significant as the observed planes around the Milky Way and Andromeda, even in the baryonic runs. 
\end{abstract}

\begin{keywords}
galaxies: haloes -- galaxies: kinematics and dynamics -- galaxies: dwarf -- Galaxy: structure -- Galaxy: disc
\end{keywords}

\section{Introduction}

The 11 classical satellites of the Milky Way have been known to lie on a thin plane with polar alignment for some time now \citep{Lynden-Bell1976}.  More recently, they have been shown to exist in a rotationally coherent structure \citep[e.g.,][]{Metz2008, Metz2009, Pawlowski2013}, and newly found ultra-faint dwarfs also lie in the plane \citep{Pawlowski2015a}. Early hints of a similar plane in Andromeda \citep{Koch2006} have recently been shown to be a highly significant thin plane \citep{Conn2013} 
with 13 of 15 of Andromeda's satellites possibly rotating in the same direction \citep{Ibata2013}.  \citet{Ibata2014a} also reported the possible discovery of planar structures outside of the Local Group, with 20 out of 22 massive nearby galaxies possibly having co-rotating planes like that of Andromeda \citep[though see][]{Cautun2015, Phillips2015}.

There have been a number of investigations using N-body simulations that have tried to quantify whether satellite planes are common in Lambda Cold Dark Matter ($\Lambda$CDM) cosmology. Studies of high-resolution N-body simulations do suggest anisotropies due to filamentary accretion from the cosmic web \citep{Donghia2008, Li2008, Libeskind2005, Libeskind2014, Lovell2011, Goerdt2013, Tempel2015, Buck2015b}. 
Filamentary infall is generally found to lead to alignment with the parent halo's angular momentum, preferentially orienting the satellites along the outer halo's major axis.  Assuming the inner and outer halo are aligned, and the disc lies along the halo's major axis, this orientation would lead to satellites preferentially orbiting in the same plane as the disc.   

While this orientation seems to be in agreement with observations of relatively massive, red, spheroidal galaxies \citep{Brainerd2005, Yang2006, Bailin2008, Welker2015}, it is at odds with the known satellite planes around the Milky Way, where the plane is almost perpendicular with the Galactic disc, and Andromeda, whose plane is is tilted $\sim 38^\circ$ from the disc \citep{Ibata2013}.  In simulations where polar planes of satellites are found around disc galaxies, the disc is oriented with the minor axis of the halo \citep{Libeskind2007}.  Non-polar planes like in Andromeda may indicate that the angular momentum of the inner and outer dark matter halo are not aligned \citep{Faltenbacher2007, Deason2011, Shao2016}.

Despite the fact that planar satellite distributions have been found in $\Lambda$CDM simulations   
\citep{Libeskind2009, Deason2011, Gillet2015, Buck2015, Cautun2015a, Sawala2016}, there is controversy regarding whether these planes resemble those found around the Milky Way and Andromeda.  In particular, the thinness of the observed planes and the number of apparent co-rotating satellites are rarely as significant in simulations as observed \citep{Pawlowski2014}. From their measurements, \citet{Ibata2013} claim a low likelihood of the satellite plane of Andromeda forming by chance. 
Conflicting conclusions have been drawn from satellite arrangements in the same simulations studied by different authors.  
While multiple authors have claimed that vast, thin planes like the one observed around Andromeda are common in the Millenium-II simulation \citep{Wang2013, Bahl2014}, others \citep{Ibata2014, Pawlowski2014} claim from the same simulation that Andromeda-like planes are very rare.  

Nearly all of the simulation studies to date on planar satellites have made use of dark matter-only simulations.  \citet{Sawala2014} claimed to find prominent planes in baryonic simulations of Milky Way-mass galaxies, but did not compare to a dark matter-only version to see if baryonic physics makes planes more prominent \citep[also, their claim of significant planes has been refuted using the same data by][]{Pawlowski2015}.  There are multiple reasons why we might expect the baryonic satellite distribution to be different than in an exact same simulation using dark matter-only.  For example, many authors have found that satellites that may survive in a dark matter-only run are completely destroyed in a baryonic run by the presence of the disc \citep[e.g.,][]{Donghia2010, RomanoDiaz2010, Zolotov2012, Brooks2014, Wetzel2016}. Additionally, \citet{Read2009} found that the presence of a disc preferentially dragged massive merging satellites into the disc plane, where they are tidally destroyed.  Might these trends somehow tend to leave planes?  In particular, would the effect of a disc preferentially destroy in-plane satellites, and be more likely to leave polar planes?

To address these questions, in this paper we look at four high-resolution ``zoom-in'' simulations of Milky Way-mass galaxies, run both dark matter-only and with baryons, in order to investigate the existence and formation of significant planes. 
This paper is organized as follows: Section \ref{SimulationData} presents the details of our simulations and dataset while section \ref{IdentifyingSatellites} explains how our luminous satellite population was chosen.  
Section \ref{GeneralObservations} discusses the orbital distribution of the entire subhalo population (not just luminous satellites). This is followed by Section \ref{PlaneDetection}, where we detail our plane detection method, and the quantitative significance of the resulting planes in Section \ref{Significance}. Section \ref{Details} discusses the characteristics of individual planes, and we examine the role of filamentary accretion in creating coherently rotating planes in Section \ref{Filaments}. In Section \ref{Comparison} we demonstrate that baryons create different planes than found in dark matter-only runs.  Finally, we summarize in Section \ref{Conclusion}.

\section{Simulation Data} \label{SimulationData}
The simulations used in this paper were run at high-resolution using the N-body + SPH (smoothed particle hydrodynamics) code \textsc{Gasoline} \citep{Wadsley2004}. The four Milky Way mass ($\sim 10^{12} \mathrm{M}_{\odot}$) haloes that are used in this paper were selected from a uniform resolution, dark matter-only box, 50 comoving Mpc on each side. The initial conditions for this box used a WMAP Year 3 cosmology (Spergel et al. 2007) with $\Omega_{m}$ = 0.24, $\Omega_{\Lambda}$ = 0.76, $H_0$ = 73 km s$^{-1}$, and $\sigma_8$ = 0.77. The work here looks only at  zoomed-in regions of these four haloes which were resimulated at higher resolution both with and without baryons \citep{Katz1993}. During the resimulation, the highest resolution particles are introduced to the region that ends up  within several virial radii of the selected halo while the rest of the box is kept at low resolution.  Maintaining the 50 Mpc box allows for large-scale transfer of angular momentum. The force resolution for the high resolution region is 173 pc. The high-resolution dark matter particles have masses of $1.3 \times 10^5\ \mathrm{M}_{\odot}$, while the gas particles start with $2.7 \times 10^4\ \mathrm{M}_{\odot}$. 
\citep{Shen2010} and a model of H$_{2}$ creation and destruction by Lyman-Werner radiation, and shielding of HI and H$_2$ described in \citet{Christensen2012}. To simulate reionization, a uniform UV background turns on at $z$ = 9 \citep{Haardt2001}. For the sub-grid feedback model, 10$^{51}$ erg of thermal energy is deposited by Type II supernovae into the surrounding gas and cooling is disabled for a period of time equal to the momentum conserving phase of the blastwave \citep{Stinson2006}. 

The overall merger histories of the four haloes in this study are quite different from each other. The heaviest, h239, has a busy merger history full of small mergers spread out in time. h258 has an almost 1:1 co-rotating merger at around z=1 which leads it to have a measurable dark disc \citep{Governato2009,Read2009}. h277, which is the closest analog to the Milky Way \citep{Loebman2014} has a very quiescent life while h285 has a violent merger history with multiple simultaneous mergers. It also has a counter-rotating major merger between z=0.8 and z=1.4 which leads it to have counter-rotating dark matter in its inner regions \citep{Sloane2016}. Halo masses\footnote{As discussed in \citet{Munshi2013} and \citet{Sawala2012}, SPH halo masses are generally lower than the same halo in a dark matter-only run. At Milky Way masses, this is primarily due to the fact that feedback removes material, and thus fitting to the same overdensity leads to a slightly smaller virial radius, as can be seen in Table \ref{table:parents}.  For the one galaxy in Table \ref{table:parents} in which the SPH run appears more massive, it is due to infalling substructure in the SPH case that isn't yet infalling in the dark matter-only run.} and virial radii for the four dark matter-only simulations and their SPH counterparts are listed in Table \ref{table:parents}.

\begin{table}
\centering
\caption{Properties of parent haloes}
\begin{tabular}{lcc}
\hline

Galaxy & $m_{vir}$ ($10^{12}\ \mathrm{M}_\odot$) & $r_{vir}$ (kpc)\\
\hline
h239 & 0.930 & 253 \\
h239+SPH & 0.924 & 252 \\
h258 & 0.817 & 242 \\
h258+SPH & 0.780 & 238 \\
h277 & 0.748 & 235 \\
h277+SPH & 0.695 & 230 \\
h285 & 0.726 & 233 \\
h285+SPH & 0.935 & 253 \\
\hline

\label{table:parents}
\end{tabular}
\end{table}

\begin{figure}
\begin{center}
\includegraphics[width=0.48\textwidth]{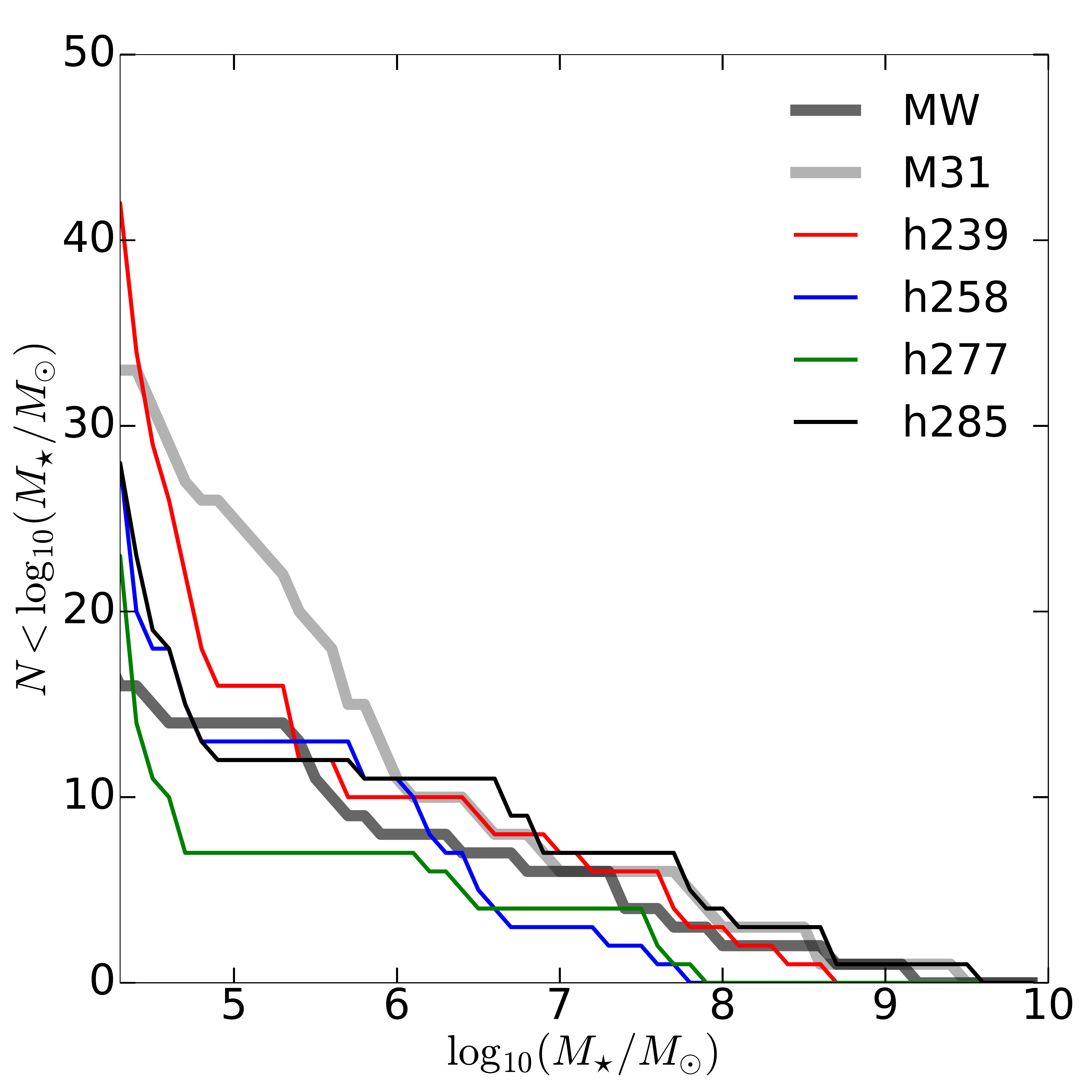}
\end{center}
\caption{Stellar mass function of the luminous satellite population around our four DM+SPH galaxies compared to those around the Milky Way and Andromeda. Milky Way data collected from \citet{McConnachie2012} and \citet{Bechtol2015}. Lower cutoff of x-axis is a result of our selection criterion of subhaloes that contain more than $2\times10^{4}\ \mathrm{M}_{\odot}$ in stellar mass at $z=0$ (more than 3 star particles).}
\label{fig:fig_MF}
\end{figure}

Two of these four galaxies (h258 and h277) were studied by \citet{Zolotov2012} and \citet{Brooks2014}, who showed that the resulting satellite luminosity functions are in good agreement with those for the Milky Way and Andromeda. They were also the first Milky Way-mass simulations to simultaneously reproduce both the luminosities and velocities of satellite populations seen around the Milky Way and Andromeda. In Fig.~\ref{fig:fig_MF}, we show the satellite stellar mass function of our four baryonic simulations compared to the observed satellite stellar mass functions for the Milky Way and Andromeda and all are relatively good matches.

\begin{figure*}
\begin{center}
\includegraphics[width=0.95\textwidth]{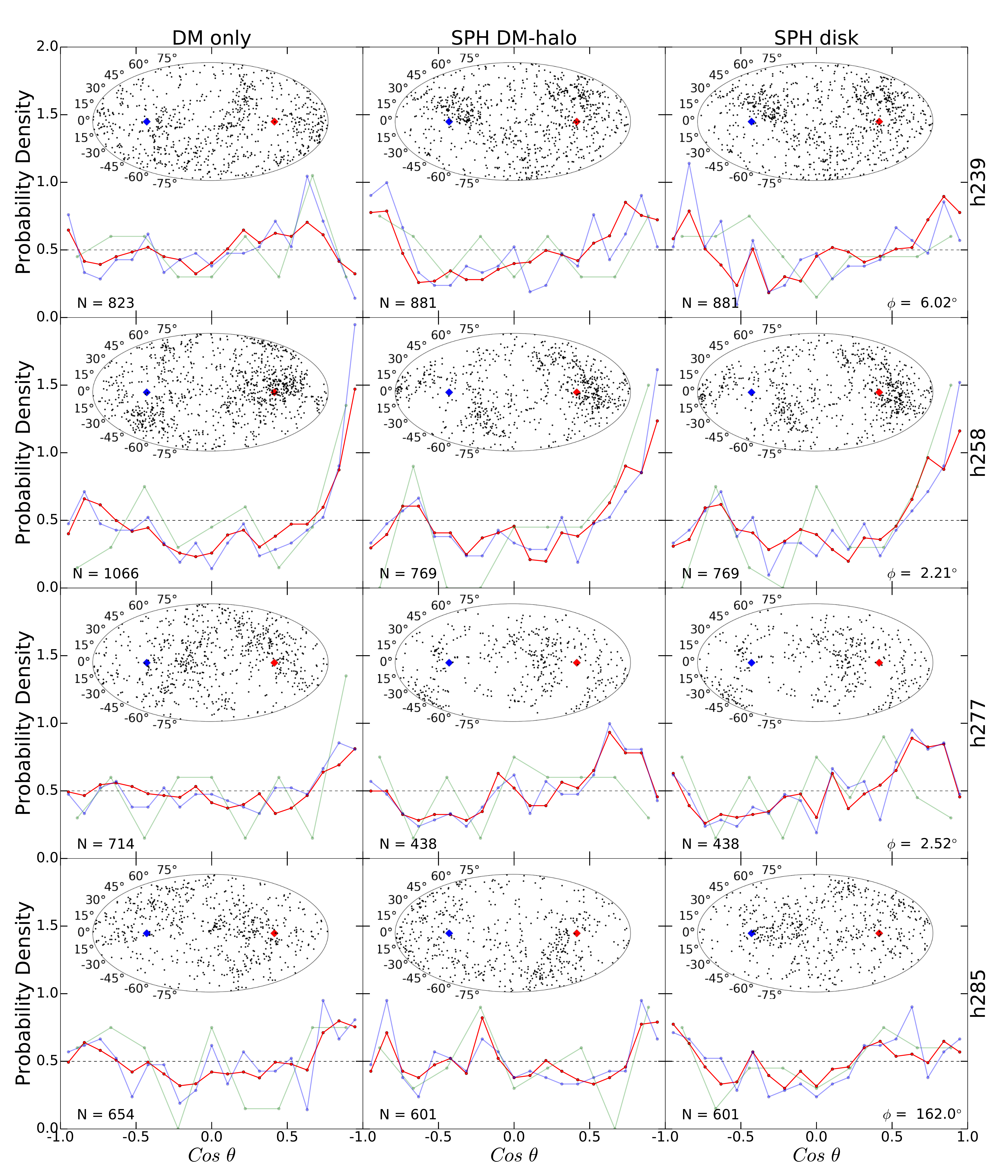}
\end{center}
\caption{Hammer-Aitoff projections of the subhalo angular momentum vectors compared to that of their parent halo. The lines in each panel show the probability distribution function of $Cos\ \theta$, where $\theta$ is the angle between the angular momentum vector of the orbit of a subhalo and the angular momentum vector of the parent halo. The red line is the histogram of the whole sample of $N$ subhaloes while blue and green are the 200 and 30 most massive subhaloes at z=0, respectively. Column 1 is with respect to the angular momentum of the dark matter halo in the dark matter-only runs, while columns 2 and 3 are the dark matter halo angular momentum and gas disc ($<$5 kpc) angular momentum, respectively, of the SPH runs. The actual spatial clustering of the tips of the subhalo angular momenta vectors are shown in the insets, with the red diamond corresponding to the parent halo angular momentum and the blue diamond representing the opposite of that vector. Generally, the subhaloes tend to rotate either aligned or anti-aligned with the halo's angular momentum. However, the trend is weak in all simulations except h258 where the subhaloes tend to heavily favor rotating with the parent halo. $\phi$ is the angle between the rotation axis of the DM halo and the disc in the SPH runs. $\phi$ is small in all cases except for h285 where the gas disc is almost counter-rotating with the dark matter halo.}
\label{fig:cosPDF}
\end{figure*}

\subsection{Identifying Satellites} \label{IdentifyingSatellites}

To identify haloes and subhaloes in our simulations we use the ROCKSTAR Phase-Space Based Halo Finder at all time steps. ROCKSTAR uses a modified friends-of-friends algorithm in six dimensions which allows for a better tracking of substructure as it interacts with its parent halo \citep{Behroozi2013}. Halo and subhalo data from ROCKSTAR is then fed through the merger tree and halo catalog generator Consistent Trees \citep{Behroozi2013} to create complete merger trees.  
A lower limit of 64 dark matter particles is imposed on the data since below 64 particles the mass function fails to converge \citep{Brooks2007}. This gives us confidence on the physical properties of haloes and subhaloes down to a virial mass of $\sim 10^{7}\ \mathrm{M}_{\odot}$. All subhaloes that exist within $1.5\times$the virial radius, ($r_{vir}$, listed in Table \ref{table:parents}), of the parent halo at $z=0$ are considered to be part of the system. The virial radius corresponds to the virial overdensity definition from \citet{Bryan1998} which is equal to 360 times the background density at z=0, but evolves with redshift.

In order to identify planes of satellites, we must first identify luminous satellites.  In the baryonic run, this is easily done using subhaloes that contain stars.  However, a different method must be used for the dark matter-only simulations.  Satellites are more commonly destroyed in baryonic runs due to the presence of a disc \citep{Donghia2010, Zolotov2012}, so it is not possible to simply identify all of the same subhaloes in the baryonic run that were found in the dark matter-only run.  Moreover, we wish to evaluate whether those studies that use dark matter-only simulations would pick out different planes than in observations or studies that use baryonic simulations. Hence, for the dark matter-only runs we use a commonly-employed method to identify the subhaloes that likely contain the most luminous satellites \citep{Ibata2014, Gillet2015}, and make no use of the fact that we know which haloes have luminous, surviving satellite counterparts in the baryonic runs.  We employ the following methods to pick out subhaloes that would correspond to luminous satellites at $z=0$ in the Universe:

$\bullet$ {\textit{Dark matter-only simulations}}:
From the surviving subhaloes at $z=0$, we pick out the 30 that were most massive at infall, under the assumption that these correspond to the 30 most luminous satellites at $z=0$.  Infall is defined as when the virial radius of the subhalo intersects with the virial radius of the parent halo for the first time.  We chose 30 subhaloes because this provides a similar number to the observed satellite counts in both the Milky Way and Andromeda (see Fig.~\ref{fig:fig_MF}).

$\bullet$ {\textit{Dark matter + SPH simulations}}:
We select subhaloes that contain more than $2\times10^{4}\ \mathrm{M}_{\odot}$ in stellar mass at $z=0$ (a minimum of 4 star particles), because this picks out approximately 25-35 subhaloes for each of our galaxies (i.e., comparable numbers to the known luminous satellites in M31 or the Milky Way).  This stellar mass lower limit also ensures that ultra-faint satellites are included in the luminous sample, since ultra-faints are members of the observed planes in M31 and the Milky Way.  We verified that selecting all satellites with $V$-band magnitude brighter than $-5$ provides a sample consistent with our stellar mass selected sample. Note that our selection may include subhaloes that have been substantially tidally stripped but still have a significant stellar mass, so that they are not necessarily the most massive satellites at $z=0$, but are the most luminous.  

Table \ref{table:results} shows the total number of subhaloes and selected luminous satellites in our simulations, $N_{subhaloes}$ and $N_{sats}$, compared to the known values in the Milky Way and Andromeda. In Appendix \ref{Appendix}, we discuss the effects of employing different selection criteria than listed above.  However, none of our conclusions below are altered if we use different selection criteria.

\begin{table*}

\caption{Plane statistics, with $\Delta$ = 20 kpc, $N_{planes}$ = 20000 and $N_{sims}$ = 2000. Column 1 listes the name of the galaxy simulation. Column 2 lists the total number of subhaloes in the simulation that are within 1.5 $r_{vir}$. Column 3 lists the number of satellites that were selected in our ``observational'' sample. For the dark matter-only runs, this is the top 30 by mass at infall. For the baryonic runs, this includes all subhaloes with $M_{star}>2\times10^4\ \mathrm{M}_{\odot}$ at $z=0$. Columns 4 and 5 list the number of subhaloes in the maximum plane and the subset of those that are co-rotating, respectively. Columns 6, 7 and 8 are the positional p-value, kinetic p-value and their product, the total p-value of the maximum plane, respectively. Columns 9 and 10 are the average r.m.s. thickness and radial extent of the maximum plane with 2$\sigma$ standard errors around the mean. Column 11 is the angle between the plane's rotation vector and the dark matter halo angular momentum vector for the dark matter-only runs or the inner gas disc angular momentum vector for the SPH runs.  Milky Way and Andromeda numbers collected from \citet{Ibata2013}, \citet{Gillet2015}, \citet{Pawlowski2015a} and \citet{Torrealba2016}.}
\begin{tabular}{lcccccccccc}
\hline
Galaxy & $N_{subhaloes}$ & $N_{sats}$ & $N_{max}$ & $N_{cor}$ & $p_{pos} (\%)$ & $p_{kin} (\%)$ & $p_{tot} (\%)$ & $\sigma_\perp (kpc)$ & $\sigma_\parallel (kpc)$ & $\psi$ \\
(1) & (2) & (3) & (4) & (5) & (6) & (7) & (8) & (9) & (10) & (11) \\

\hline
\hline

h239 & 823 & 30 & 13 & 8 & 7.55 & 58.1 & 4.39 & 14.0$\pm$2.8 & 200.8$\pm$53.6 & $62.7^\circ$ \\
h239+SPH & 881 & 42 & 15 & 8 & 3.45 & 100.0 & 3.45 & 11.6$\pm$3.4 & 197.2$\pm$52.9 & $37.8^\circ$ \\
h258 & 1066 & 30 & 10 & 8 & 71.5 & 10.9 & 7.81 & 12.9$\pm$3.4 & 176.7$\pm$45.4 & $35.0^\circ$\\
h258+SPH & 769 & 28 & 10 & 7 & 69.9 & 34.4 & 24.0 & 13.6$\pm$3.7 & 184.8$\pm$51.2 & $17.1^\circ$\\
h277 & 714 & 30 & 12 & 6 & 57.9 & 100.0 & 57.9 & 10.1$\pm$3.0 & 165.8$\pm$47.4 & $41.9^\circ$\\
h277+SPH & 438 & 23 & 10 & 6 & 8.75 & 75.4 & 6.60 & 9.7$\pm$2.7 & 197.3$\pm$53.9 & $65.7^\circ$\\
h285 & 654 & 30 & 15 & 8 & 31.4 & 100 & 31.4 & 11.9$\pm$3.4 & 89.9$\pm$26.6  & $12.2^\circ$\\
h285+SPH & 601 & 28 & 10 & 8 & 68.2 & 10.9 & 7.46 & 8.6$\pm$2.5 & 154.4$\pm$53.6 & $31.7^\circ$\\

\\
\hline
\\

MW &  & 48 & 11 & 8 & & & &\\
M31 (Gillet2015) & & 27 & 14 & 12 & 1.60 & 1.3 & 0.0208 & 12.5 & 154.7\\
M31 (Ibata2013) & & 27 & 15 & 13 & 0.13 & 0.74 & 0.00096 & 12.6 & 191.9\\
\hline
\end{tabular}
\label{table:results}

\end{table*}

\section{Orbital Properties of All Subhaloes} \label{GeneralObservations}

Before we examine the distribution of only these most massive/luminous haloes, we first look at the general characteristics of all $N_{subhaloes}$ in our sample for each simulated galaxy (all subhaloes with $\geq$ 64 particles within $1.5\times r_{vir}$ of the parent halo). Orbital information for all subhaloes is shown in Fig.~\ref{fig:cosPDF}. For each of our galaxies, the colored lines in column 1 shows the angular momentum vectors of the subhalo orbits with respect to the parent halo's angular momentum in the dark matter-only simulation. The colored lines in columns 2 and 3 show the same information with respect to either the dark matter angular momentum or gas disc angular momentum, respectively, in the baryonic simulation. The dark matter angular momentum vectors of the parent halo are calculated by summing the angular momenta of all dark matter particles within the virial radius, while the gas disc angular momentum vector is calculated similarly by using all gas particles within a 5 kpc radius. 

The insets in each panel of Fig.~\ref{fig:cosPDF} show the actual spatial clustering of the tips of the subhalo angular momentum vectors in a Hammer-Aitoff projection. The red and blue diamonds represent the parent halo (or disc) angular momentum vector and its opposite, respectively.  The angle between the parent halo's dark matter halo and disc angular momentum vectors in the SPH simulations is noted as $\phi$ in column 3. 

$\theta$ is the angle between the angular momentum vector of a subhalo's orbit and the angular momentum vector of the parent halo. Therefore $Cos\ \theta$ ranges from -1, where the subhalo orbit is completely opposite to the average rotation of the parent halo (or disc), through 0, where the subhalo orbit is perpendicular to the rotation of the parent, to 1, where the subhalo rotation lines up with the parent halo (or disc) rotation. The probability density graph which shows the distribution of subhaloes with a certain value of $Cos\ \theta$ is represented by three lines: red, blue, and green corresponding to the whole population, most massive 200, and most massive 30 subhaloes at $z=0$, respectively. This breakdown helps distinguish any mass dependent properties of subhalo orbit orientation, but none is seen.

Generally, the subhaloes have a slight tendency to rotate either aligned or anti-aligned with the halo's angular momentum. However, the trend is weak in all simulations except h258 where the subhaloes tend to heavily favor rotating with the parent halo, as seen in the clustering of points around the red parent angular momentum vector.  Because the disc and halo angular momentum vectors are also nearly aligned in h258, the subhaloes also are aligned with the angular momentum vector of the disc.  
The angular momentum vector of the disc of h285 is almost completely anti-aligned with the dark matter halo's angular momentum in the SPH run (by $162^\circ$). This results in the subhalo alignment trend being reversed when compared to the dark matter halo angular momentum vector versus the disc vector in the SPH run. This scenario is a result of a large counter-rotating merger in the history of h285. The alignment trend in the other three galaxies is the strongest when every subhalo in our sample is selected.  Selecting for the most luminous satellites still keeps the general trend, albeit more weakly.

\section{Plane Detection} \label{PlaneDetection}

\subsection{Plane detection algorithm}

The plane detection method used in this work follows the procedure used by \citet{Gillet2015} and subsequently \citet{Buck2015}. Twenty thousand random planes, defined by their normal vectors, all passing through the centre of the parent halo, are generated at $z=0$. The number of subhaloes that fall within a distance $\Delta = 20$ kpc of these planes are recorded and are considered part of the plane. Just as in \citet{Gillet2015}, $\Delta$ is chosen to be roughly 3$\times$ the root mean square (r.m.s.) thickness of the M31 plane to capture planes of similar thickness in our simulations (which range from 8.6 to 14.0 kpc, comparable to the 12.5 kpc of Andromeda, see Table \ref{table:results}). The plane with the highest number of subhaloes is termed the maximum plane, and the number of subhaloes in that plane is labelled $N_{max}$. This is the plane that is selected and its kinematics are studied. The algorithm is run multiple times on each parent halo to make sure that the number of random planes generated is enough to pick out the same max plane on every iteration. 

If more than one plane is detected with the same maximum number of subhaloes, the one with the most co-rotating subhaloes (largest $N_{cor}$) is chosen as the maximum plane. 

Co-rotation is a binary value with subhaloes either rotating with or against the plane depending on which side their angular momentum vector falls. 

The plane detection method leads to unique and robust planes. Planes with 1 or 2 fewer subhaloes than the maximum plane always turn out to be the maximum plane with a few members missing because the test plane used to detect this second plane was slightly offset from the one that detected the maximum plane. The next plane down from the maximum plane always has significantly fewer subhaloes, with one exception\footnote{h277 dark matter-only had another plane with the same $N_{max}$ but smaller $N_{cor}$}. Thus, in general there is no ambiguity in what the maximum plane is for a given simulation.

\subsection{Significance of detected planes} \label{Significance}

\begin{figure*}[h]
\begin{center}
\includegraphics[width=0.9\textwidth]{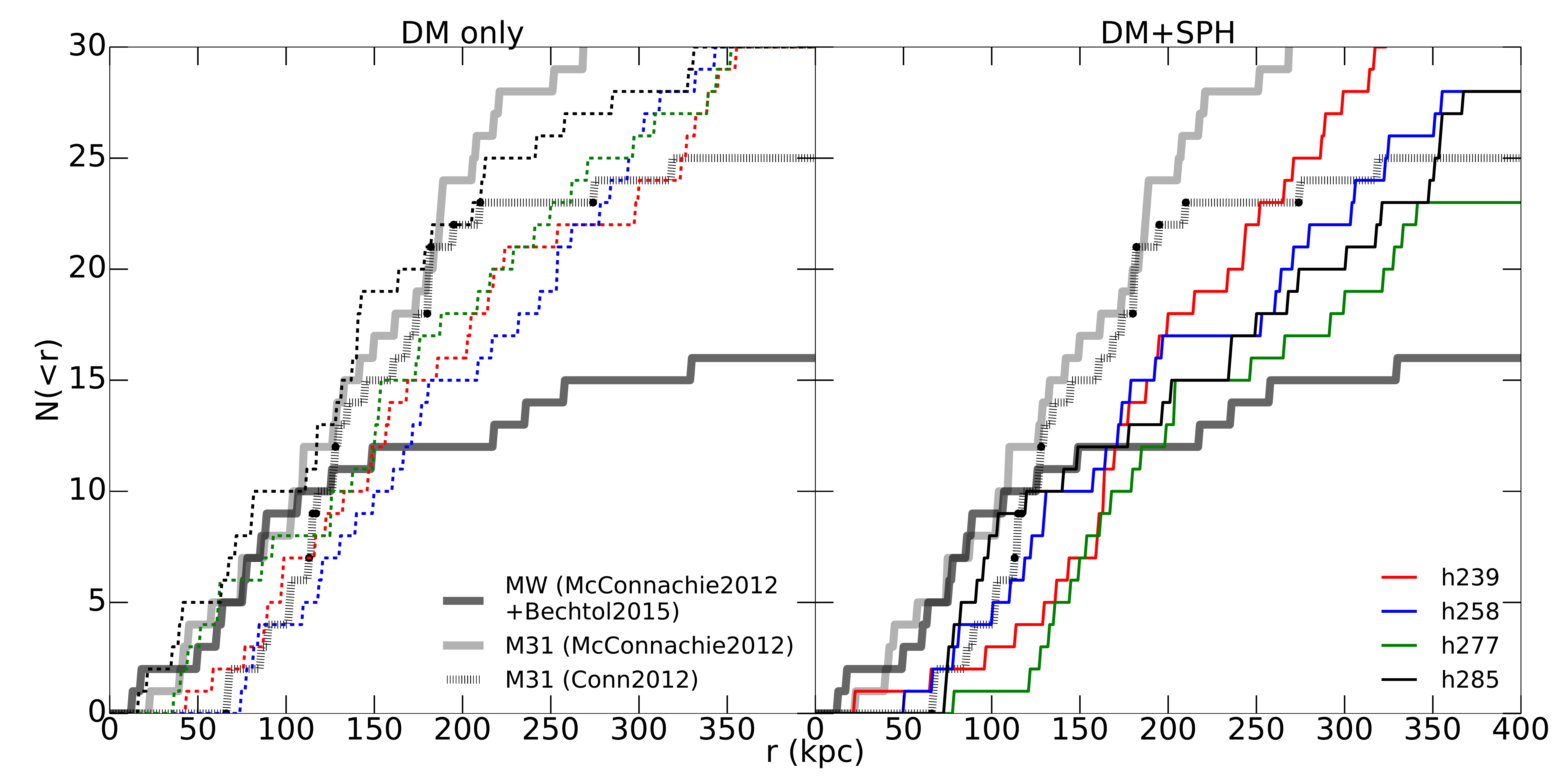}
\end{center}
\caption{Cumulative radial distribution of satellites using our selection methods at z=0 (30 most massive at infall in the dark matter-only runs, all satellites with M$_{star} > 2\times10^4\ \mathrm{M_{\odot}}$ in the SPH runs). Colors represent different parent haloes; solid lines are baryonic runs while dashed lines are dark matter-only runs. Milky Way and Andromeda satellites with M$_{star} > 2\times10^4\ \mathrm{M_{\odot}}$ are shown in thicker black and gray bold lines, respectively \citep{McConnachie2012,Bechtol2015}. Grey dotted line shows the Andromeda satellites from \citet{Conn2012} as used by \citet{Gillet2015} in their analysis. In general, the SPH simulations have a less centrally concentrated population of subhaloes.} 
\label{fig:cumulative_subs}
\end{figure*}

 We now proceed to quantify how statistically significant these planes are against a random distribution. To generate a random distribution of maximum planes, positional information for test subhalo populations are created conforming to the radial distribution of subhaloes present in each of the parent haloes.  The satellite radial distribution for our cosmologically simulated galaxies is seen in Fig.~\ref{fig:cumulative_subs}. The SPH runs have a less centrally concentrated satellite population when compared to their corresponding dark matter-only runs. This can be attributed to the fact that subhaloes in the SPH runs have a higher chance of being destroyed the closer they are to the disc  \citep{Donghia2010, Schewtschenko2011, Brooks2013, Brooks2014, Wetzel2016}. In general (apart from h239) not only are the SPH runs less concentrated, but the subhalo numbers are reduced overall, see Table \ref{table:results}. 

Fig.~\ref{fig:cumulative_subs} also illustrates that the cumulative radial distributions of our subhaloes in the SPH simulations qualitatively match the observations of the Milky Way and Andromeda, barring the effects of incomplete surveys that reduce the number of ``faint'' subhaloes present around the Milky Way at higher radii \citep[for a detailed discussion, see][]{Yniguez2014}. 
  
For this comparison, we use all known satellites of the Milky Way and Andromeda that have M$_{star} > 2\times10^4\ \mathrm{M_{\odot}}$ \citep{McConnachie2012, Bechtol2015}. No assumption is made about the stellar mass of the subhaloes selected from the dark matter-only run, other than that the 30 most massive subhaloes at infall are likely to correspond to the 30 most luminous satellites.  
 
\begin{figure*}[h]
\begin{center}
\includegraphics[width=0.9\textwidth]{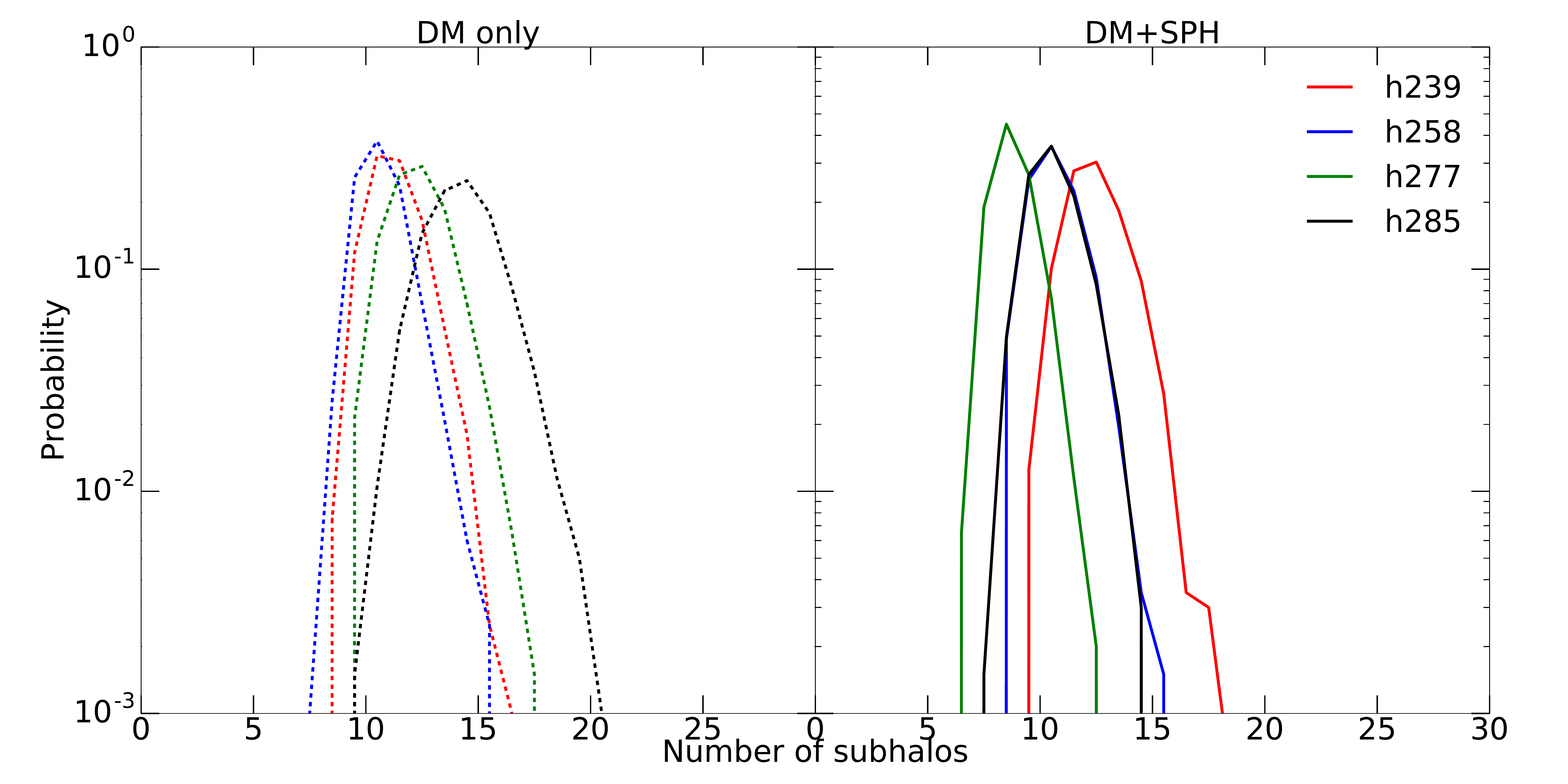}
\end{center}
\caption{Probability distribution function of the maximum plane generated from 2000 instances of subhalo positions following the radial subhalo distributions in our simulations (Figure \ref{fig:cumulative_subs}). Peak indicates the number of subhaloes most likely to be found in the maximum plane. } 
\label{fig:pdf}
\end{figure*}

Next the plane finding algorithm is run on a large number ($N_{sims} = 2000$) of test cases where $N_{sats}$ is randomly sampled from a radial distribution that matches the radial distributions of each parent halo in order to build up a statistical probability distribution function (pdf) of finding a plane with a certain number of subhaloes given that radial distribution. This is illustrated in Fig.~\ref{fig:pdf} where we can see that the most likely number of subhaloes found in the maximum plane with $\Delta$ = 20 kpc is somewhere between 10-15. 
With these pdfs we can find the probability of finding $n$ or more satellites in a plane given a population of $N$ satellites, i.e. the positional p-value, where,

\begin{equation}
p_{pos} = p(X \geq n) = \sum_{n}^{N} pdf .
\end{equation}

Since our method forces a binary choice on the rotation direction on any given subhalo (depending on whether their angular momentum vector falls either on one side of the plane or the other), we can easily also find the kinematic p-value which is the probability of finding $k$ or more plane satellites corotating given that the plane contains $n$ satellites.

\begin{equation}
p_{kin} = p(X \geq k) = 2\times \sum_{k}^{n} p(i) 
\end{equation}

with $p(i)$ being the binomial distribution:

\begin{equation}
p(i) = \binom{n}{i}\lambda^{i}(1-\lambda)^{n-i} 
\end{equation}

where $\lambda = 0.5$.

The total probability of finding $n$ out of $N$ satellites with $k$ co-rotating given a certain radial distribution is found by multiplying $p_{pos}$ and $p_{kin}$ together. The results are presented in Table \ref{table:results}.

Looking at the p-values of our galaxies in Table \ref{table:results}, the first thing to note is that there is generally no correlation in the p-values between the dark matter-only and the SPH versions, i.e. a smaller p-value in the dark matter-only simulation does not imply similar results in the corresponding SPH simulation or vice versa.   However, we show in Section \ref{Comparison} that this is because the planes picked out in the dark matter-only runs and the SPH runs are different.  This is the first effect that results from including baryons: a different set of satellites are inside the halo at $z=0$.  Because of this, the dark matter-only versions of the runs cannot be compared directly to their SPH counterparts.  In the remainder of this section, we simply treat them as additional examples of planes of satellites.

Examining the total p-values of our galaxies in Table \ref{table:results}, we find that over 50 per cent of our sample have p-values below 10 per cent.  While this value is nowhere near as significant as in the Milky Way or M31, it is much lower than the p-values found in the simulations studied by \citet{Gillet2015} using the same method (where the lowest p-value was $\sim$14 per cent).  In three of our haloes, this low p-value results from low positional p-values, and in the two others it results from low kinematic p-values.  In no case is there a simulation plane with both a low $p_{pos}$ and a low $p_{kin}$, which is the case in M31.  Because of this, none of our galaxies have planes as significant as M31.  

Included in Table \ref{table:results} are the numbers of satellites that make up the planes in the Milky Way \citep{Pawlowski2015a, Torrealba2016} and Andromeda.  We include two different estimates of p-values for Andromeda \citep{Ibata2013, Gillet2015}. Our smallest p-value is still substantially higher than the estimated p-value of the plane of satellites observed around Andromeda of 0.0208 per cent \citep{Gillet2015}. \citet{Cautun2015a} recently suggested that the significance of the satellite planes observed around the Milky Way and Andromeda may have been overestimated by around an order of magnitude because the significance of the planes are very sensitive to small changes in the sample selection criteria. However, even if the Milky Way and Andromeda planes are less significant than previously estimated, the significance of our planes is still much lower. 

The total numbers of satellites found in our planes are comparable to the Milky Way and M31 values (this is a result of choosing $\Delta=20\ \mathrm{kpc}$).  Importantly, we can identify that the reason that some galaxies have low positional p-value is due to the fact that their satellites are less radially concentrated.  For example, the baryonic versions of h239 and h277 are the least radially concentrated within $\sim$150 kpc (see Figure \ref{fig:cumulative_subs}).  Because all of our planes are made to pass through the centre of the halo and have a thickness $\Delta = 20$ kpc, as the satellite distribution becomes more centrally concentrated the number of satellites that make up the plane tends to increase.  In our random distributions with similar radial concentrations, it becomes much more difficult to pack as many satellites into the maximum plane as the radial concentration decreasees.  Hence, the significance of the haloes being in a planar structure increases compared to random if the satellites are less radially concentrated.  

This is the second effect that results from including baryons: if inclusion of baryons makes the satellite distribution less radially concentrated (as seen in Figure \ref{fig:cumulative_subs}), then {\it planes of satellites will be more significant compared to a random distribution when baryons are included.}

In fact, this is part of the reason that Andromeda has such a high positional p-value.  As an example, $N_{sats}$ and $N_{max}$ for the dark matter-only h285 run are similar to those for M31, but their $p_{pos}$ values are quite different, with our simulation planes being much less significant. The reason for this can be seen clearly in Fig.~\ref{fig:cumulative_subs}, where the satellite distribution for h285 dark matter-only is much more centrally concentrated than the satellite distribution used in \citet{Gillet2015} from \citet{Conn2012}. This makes it much easier to find planes with the same number (or higher) of satellites in our dark matter-only h285, thus lowering their significance (leading to a higher $p_{pos}$ value).  Note as well that the updated method of distance finding in \citet{Conn2012} leads to a much less centrally concentrated satellite distribution in M31 than earlier works, like those compiled in \citet{McConnachie2012}.  Using the distribution from \citet{McConnachie2012} leads to a lower $p_{pos}$ for M31.

Our less centrally concentrated satellite radial distribution is also the reason why we find lower p-values than \citet{Gillet2015}.  Although they examine multiple methods of defining satellites to comprise their maximum planes, in all but one case their resulting satellite radial distribution is more concentrated than in Andromeda.  Their least concentrated distribution is also the one that leads to the lowest p-value.  

\begin{figure*}
\begin{center}
\includegraphics[width=1.0\textwidth]{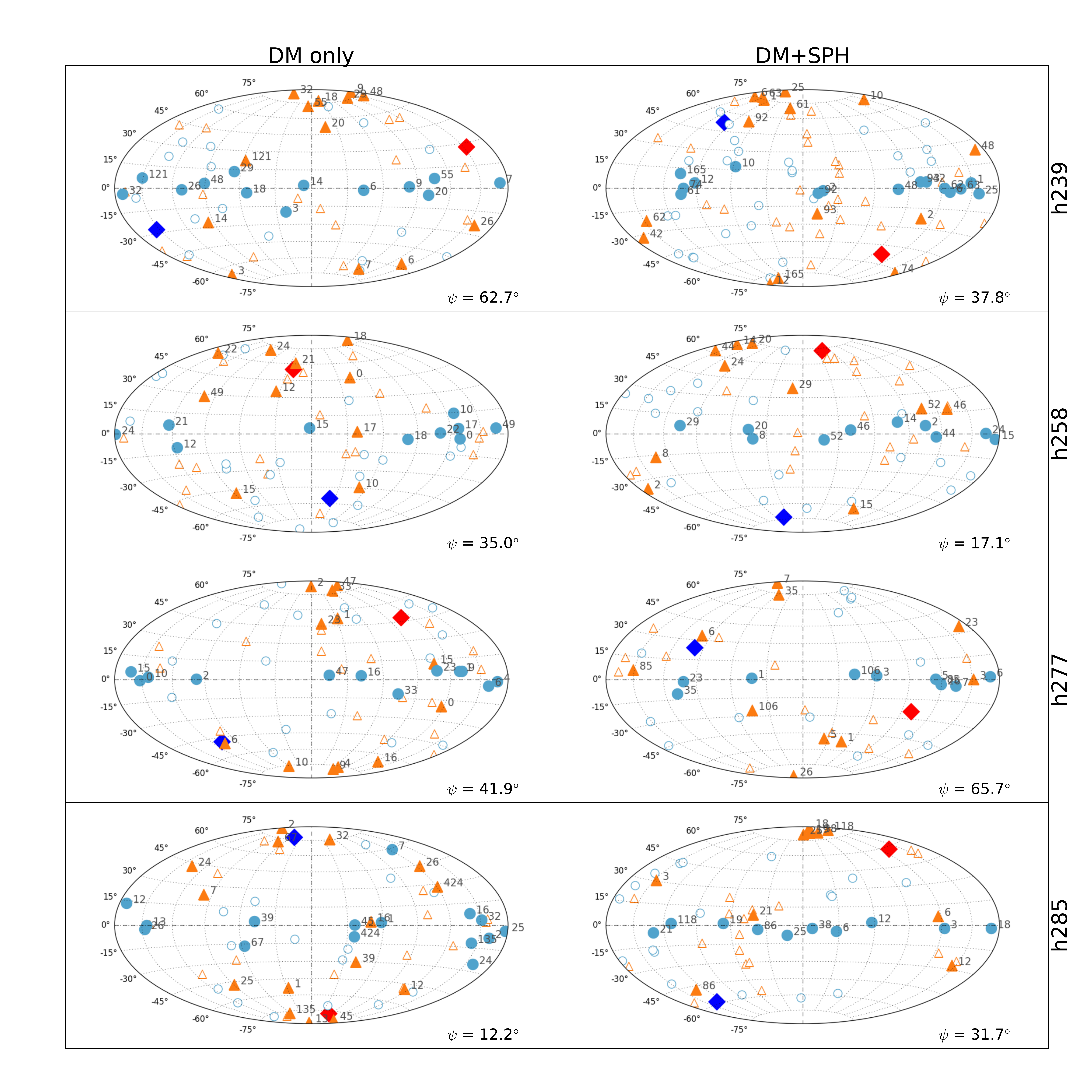}
\end{center}
\caption{Each panel shows an Aitoff-Hammer projection of the sky from the centre of our simulated galaxies. Blue circles show the position of subhaloes and the corresponding orange triangles show the tips of the angular momentum vectors of their orbit around the galactic centre projected onto the sky. The numbers are labels for the subhaloes in decreasing order of virial mass at z=0 with 0 being the heaviest. The filled points indicate the subhaloes making up the maximum plane. A clustering of the angular momentum vectors indicate that those subhaloes are coherently rotating. This can most clearly be seen in the SPH versions of h285 and h239 and in the dark matter-only versions of h239 and h277. Red (and blue) diamonds indicate the angular momentum (antiparallel-) vector of the parent halo as in Fig.~\ref{fig:cosPDF}. Note that the clustering does not usually correlate with the angular momentum vector of the parent halo, indicating that the maximum plane is usually offset from the rotation of the dark matter halo or disc. $\psi$ is the angle between the maximum plane's rotation vector and the dark matter halo angular momentum vector for the dark matter-only runs or the inner gas disc angular momentum vector for the SPH runs.}
\label{fig:planes_hammer}
\end{figure*}

Hence, it is easier for us to find relatively significant planes in terms of positional p-value.  On the other hand, it is harder for us to find significant co-rotating planes.  Our highest fraction of co-rotating satellites in a maximum plane is 80 per cent.  This is slightly higher than in the Milky Way, but not as high as in Andromeda.  Those planes with 80 per cent co-rotating satellites have the lowest $p_{kin}$ , but many of our planes have $\sim$50 per cent co-rotating, significantly lower than that of Andromeda.  As we will examine in more detail in Sections \ref{Details} and \ref{Filaments}, and in Figure \ref{fig:planes_hammer}, our ability to achieve relatively high significance in a few cases is only due to our binary definition of co-rotating.  In reality, we never achieve a true fraction of co-rotating satellites higher than $\sim$50 per cent.  Although we only have a binary definition in M31, in the Milky Way proper motion data has been used to show that 8 of the 11 satellites in the plane seem to be truly co-rotating \citep{Pawlowski2015a}.  Hence, in no case do we produce any satellite plane that is as coherently rotating as that seen in the Milky Way.

\section{Coherency of Planes}
\subsection{Details of  Individual Planes} \label{Details}

Fig.~\ref{fig:planes_hammer} shows us the position and angular momentum vectors of the most luminous and maximum plane subhaloes in our simulations as projected onto the night sky in an Aitoff-Hammer projection as seen from the centres of the galaxies. The projection plots are oriented such that (1) the maximum planes lie along the equator and (2) the angular momentum vectors of subhaloes that move along the plane cluster near the poles.  The subhaloes are labelled in decreasing order of their virial mass at z=0 out of the whole population. Each blue circle shows the position of the selected subhaloes that are being analysed for that simulation and the corresponding orange triangles show where the tips of the angular momentum vectors of their orbit around the galactic centre project onto the sky. The filled points indicate the subhaloes making up the maximum plane and only these are labelled. The angle between the plane's rotation axis and the dark matter angular momentum axis for the dark matter-only simulations or the inner gas disc angular momentum axis for the SPH simulations is listed as $\psi$ in Table \ref{table:results}.

Looking at the make-up of the maximum plane for each simulation gives us insight into their formation process. For the dark matter-only version of h239, both subhalo pairs 29 \& 48 and 18 \& 55 fall into the parent halo together with the same trajectories. However, the former pair stay relatively together throughout the simulation and end up in close proximity in the projection view in Fig.~\ref{fig:planes_hammer}. Due to slight differences in orbital radii and velocities, subhaloes 18 and 55, even though they keep similar trajectories till $z=0$, end up away from each other in proximity. The orbital coherence of both pairs can be seen in the clustering of their angular momentum vectors in Fig.~\ref{fig:planes_hammer}. The rest of the maximum plane is split evenly between subhaloes that are in coherent orbits with these two pairs but had their own individual infall trajectories and other subhaloes that are crossing the plane by chance at $z=0$. 

For the baryonic version of h239, a cluster of subhaloes (1, 6, 25, 62 \& 63) falls into the parent halo and ends up with similar positions (projected and real) and trajectories at z=0. These subhaloes make up the bulk of the maximum plane and, barring subhalo 6, also the coherently rotating part of the maximum plane. Four of these satellites are accreted close to $z=0$.  Subhaloes 1 and 63 have infall times of 12.9 and 13.5 Gyr, respectively, while subhaloes 6 and 25 are just about to come into the virial radius of the parent halo as the simulation ends at $z=0$. Subhalo 62, while part of the group, has an early infall time of 5.6 Gyr but is actually a ``splashback'' subhalo coming back from its apocentre \citep{Gill2005, Ludlow2009, Wang2009, Teyssier2012}. Subhaloes 2 and 92, though having similar trajectories and positions, have a velocity vector directed out of the plane and are only transient parts of the plane. Subhaloes 10 \& 12 also fall in together $\sim12\ \mathrm{Gyr}$ following a distinct accretion filament. These filaments are discussed in greater detail in Section~\ref{Filaments}.  

Note that, despite the fact that both the dark matter-only and baryonic versions of h239 have pairs of subhaloes falling in together to make up the maximum planes, {\it they are not the same subhaloes}.  We discuss in Section \ref{Comparison} that there is very little overlap in the subhaloes that constitute the maximum planes in the dark matter-only and baryonic simulations, in all cases.

The dark matter-only version of h258 lacks any significant coherent portion of the maximum plane and is mostly made up of rogue subhaloes that form a transient plane at $z=0$. For the baryonic version of h258, only subhaloes 44 and 46 fall in together, but due to the latter's close encounter with the core of the parent halo, they end up with substantially different positions and trajectories. The coherent part of the maximum plane (subhaloes 14, 20, 24 \& 44) have their angular momentum vectors deviating slightly from the normal of the plane, indicating that they will slowly move out of the maximum plane as time progresses.

For the h277 dark matter-only halo, subhaloes 4, 6, and 10 fall in as a large sheet and mostly retain their similar positions and trajectories due to their late infall times (not yet fallen in, 11.9 Gyr and 13.7 Gyr respectively). They are counter-rotating with respect to the other set of subhaloes that are moving coherently (by chance), 2, 33 and 47. The rest are transients.
The baryonic version of h277 has a maximum plane defined by subhaloes 7 \& 35 (co-rotating) and 26 (counter-rotating).  Subhaloes 7  and 35 had very early infall times of 2.6 Gyr but keep their coherent formation (subhalo 26 infalls at 10.3 Gyr). The rest are all transients in tilted orbits. 

The dark matter-only version of h285 has a plane made mostly of transients with little or no coherence. A few clusters that showed early coherence and infall lose those properties by $z=0$. On the other hand, the baryonic version of h285 has a strongly coherent maximum plane with three early infall subhaloes, 18, 25 \& 118 (infall times of 4.2 Gyr, 3.5 Gyr and 3.5 Gyr respectively).  These three subhaloes keep their coherent rotation till $z=0$ (their relative positions spread out due to slightly different orbital radii and speeds). Along with another coherent pair (19 \& 38), half of the maximum plane of subhaloes in the baryonic h285 run have a very tightly clustered set of angular momentum vectors implying that they will hold the structure of the plane for a long time.

\begin{figure*}
\begin{center}
\includegraphics[width=1.0\textwidth]{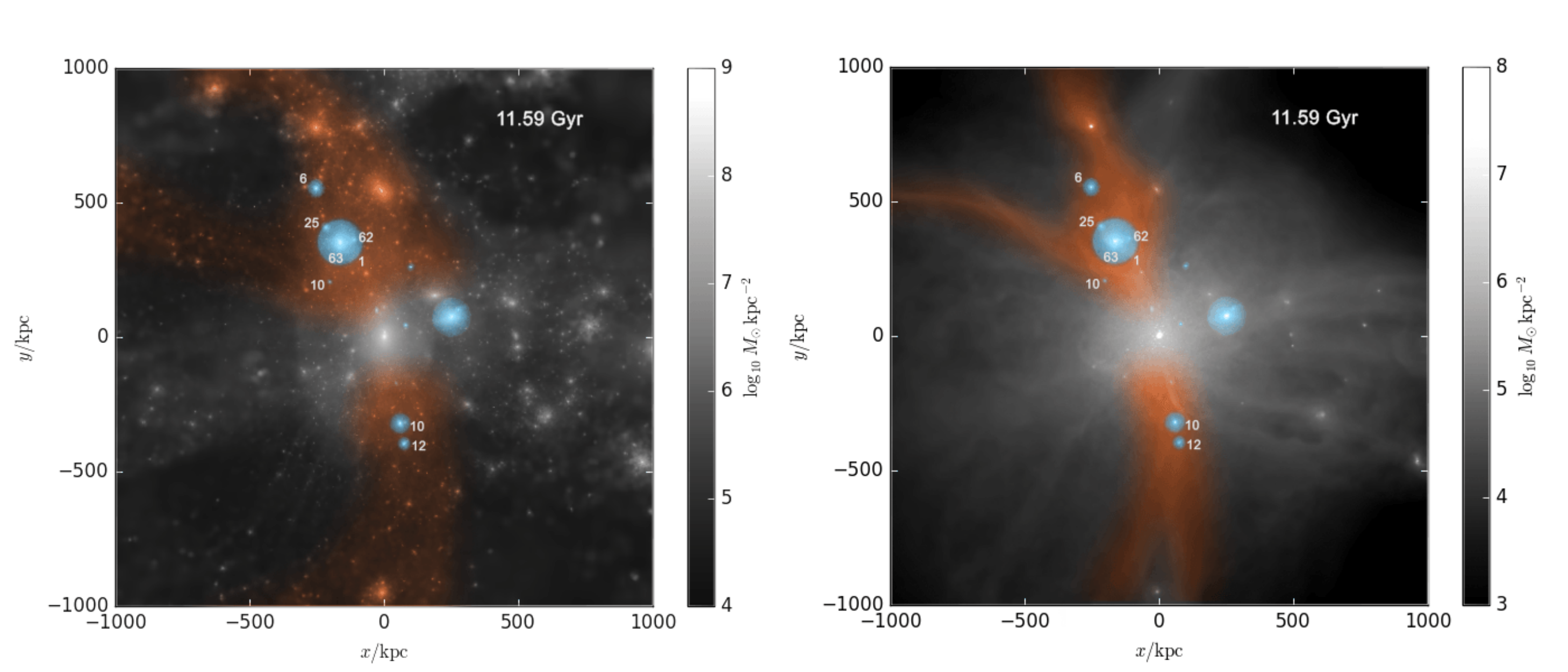}
\end{center}
\caption{Snapshots of integrated dark matter density (left panel) and integrated gas density (right panel) around the baryonic h239 simulation at $t=11.59\ \mathrm{Gyr}$. Subhaloes that will belong to the maximum plane at $z=0$ are highlighted in blue. Subhaloes mentioned in the text are labelled. Accretion filaments are highlighted in orange and distances are in physical units.  In this galaxy, the maximum plane is built by a pair of filaments that contribute most of the plane satellites. }
\label{fig:h239g_filament}
\end{figure*}

\begin{figure*}
\begin{center}
\includegraphics[width=1.0\textwidth]{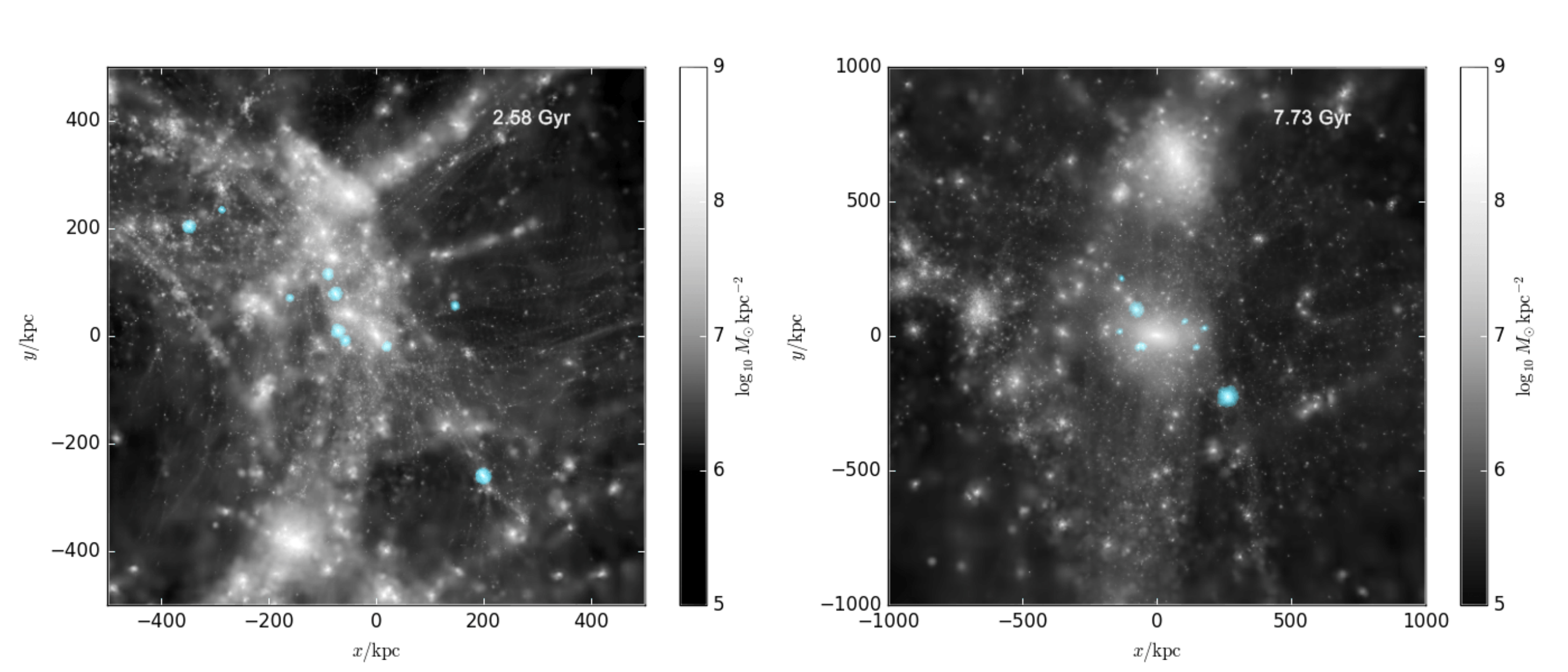}
\end{center}
\caption{Snapshots of the integrated dark matter density around h258 in the dark matter-only run at $t=2.58\ \mathrm{Gyr}$ (left panel) and $t=7.73\ \mathrm{Gyr}$ (right panel). Subhaloes that will belong to the maximum plane at $z=0$ are highlighted in blue. Distances are in physical units. No distinct filamentary accretion of subhaloes can be seen at either time.  At $z=0$, this halo has no coherently rotating plane of satellites.  }
\label{fig:fig_h258_filament_00096_00288}
\end{figure*}

\subsection{Relation to Filamentary Accretion} \label{Filaments}

Looking at the traced-back positions of the plane subhaloes, we find that groups of them that are within close proximity of each other have similar infall times and follow similar trajectories.  In this section, we investigate the role of filamentary accretion in contributing to the maximum planes, in particular the coherence of rotation in the planes.
The angular momentum vectors of the plane subhaloes confirm that many instances of these planes are transient, and are only a feature that is dependent on the time we are looking at it. This is seen clearly in Fig.~\ref{fig:planes_hammer} in every instance where a subhalo in the maximum plane does not have its angular momentum vector near either pole of the projection. Some subhaloes will keep moving coherently, but the whole plane may not exist at a different time. 

\begin{figure}
\begin{center}
\includegraphics[width=0.5\textwidth]{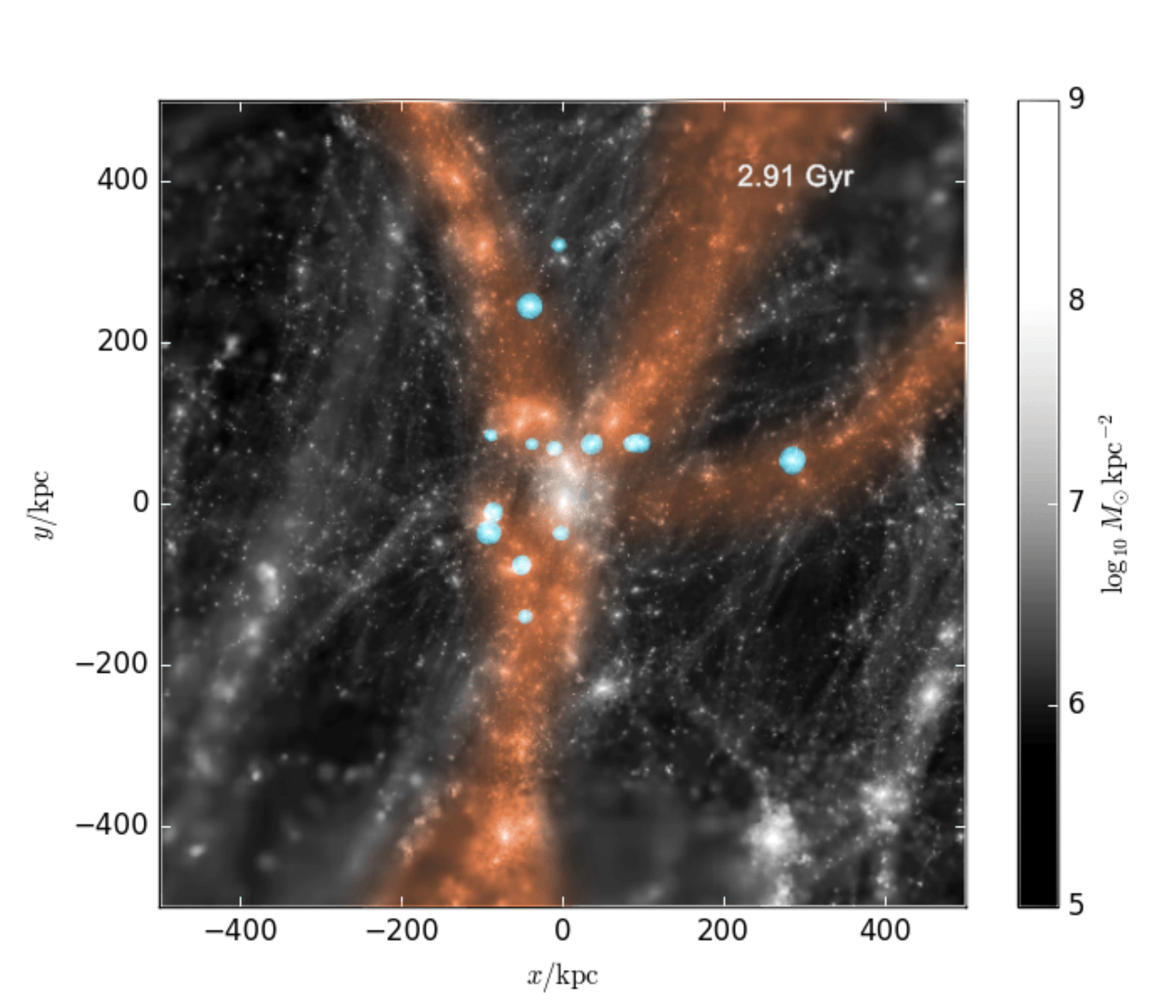}
\end{center}
\caption{Snapshot of integrated dark matter density around h285 DM at $t=2.91$ Gyr. Subhaloes that will belong to the maximum plane at $z=0$ are highlighted in blue.  Accretion filaments are highlighted in orange and distances are in physical units. While multiple distinct filaments can be seen, the maximum plane subhaloes do not have a strong affiliation with any one of them. At $z=0$, this halo has no coherently rotating plane of satellites. }
\label{fig:fig_h285_filament_00108}
\end{figure}

\begin{figure*}
\begin{center}
\includegraphics[width=1.0\textwidth]{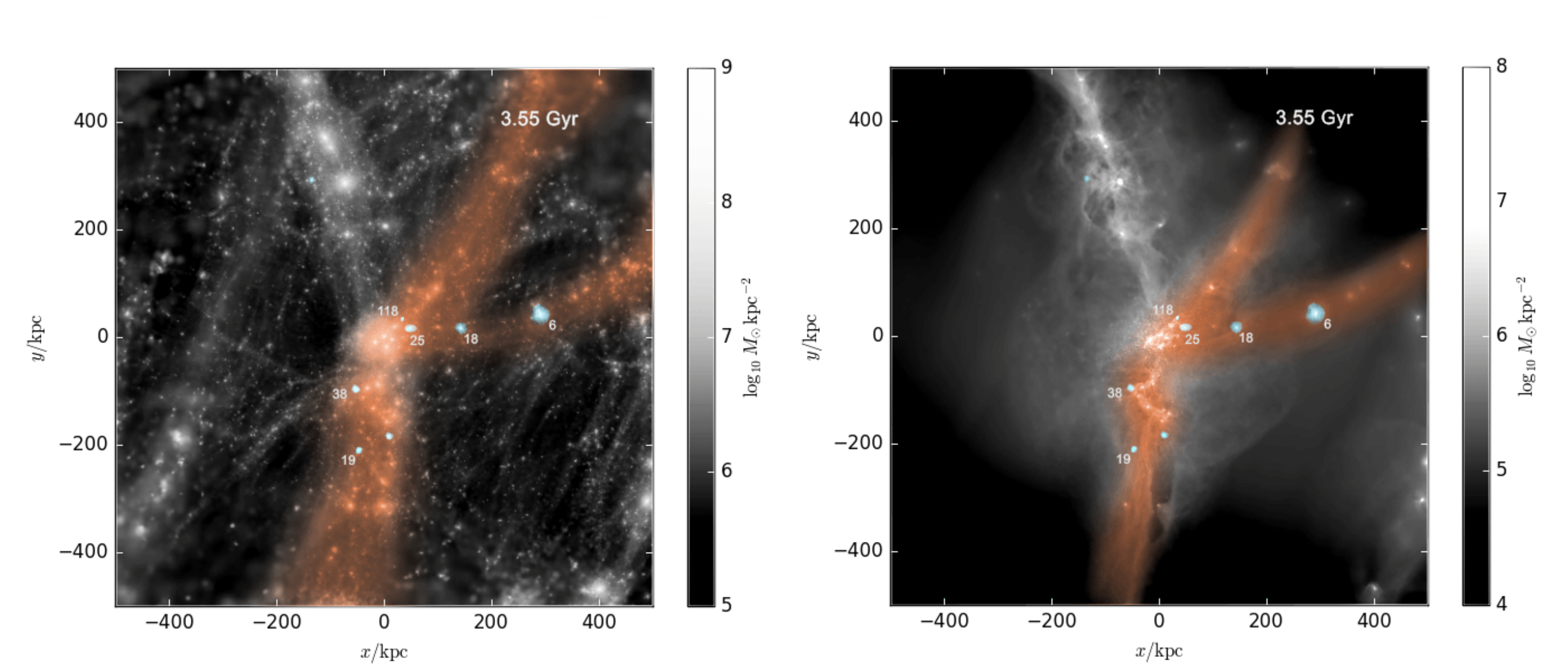}
\end{center}
\caption{Snapshots of integrated dark matter density (left panel) and gas density (right panel) around the baryonic version of h285 at $t=3.55\ \mathrm{Gyr}$. Subhaloes that will belong to the maximum plane at $z=0$ are highlighted in blue. Subhaloes mentioned in the text are labelled. Accretion filaments are highlighted in orange and distances are in physical units. The same filamentary structures as those in Fig.~\ref{fig:fig_h285_filament_00108} are seen, but in this case the maximum plane subhaloes mainly come from two filaments that are close to each other.  This configuration leads to a maximum plane at $z=0$ that is strongly coherent in its rotation.}
\label{fig:fig_h285g_filament_00132}
\end{figure*}

A good example of how filamentary accretion can lead to the formation of coherent planes is illustrated in Fig.~\ref{fig:h239g_filament}. The left and right panels are snapshots of the integrated dark matter and gas densities respectively of the SPH run of h239 at 11.59 Gyr. Recall from Fig.~\ref{fig:planes_hammer} that both versions of h239 contain maximum planes with a cluster of coherently co-rotating satellites. The orange regions  in Fig.~\ref{fig:h239g_filament} highlight the two major filaments through which most of the maximum plane subhaloes (highlighted in blue) infall. The labelled subhaloes all end up rotating coherently in the maximum plane (or in the case of subhalo 12, counterrotating). The rest of the maximum plane elements come from another distinct filament or have no association with any major filamentary structures. 

As a counter example, both versions of h258 lack a coherently rotating maximum plane at $z=0$. Fig.~\ref{fig:fig_h258_filament_00096_00288} shows the infalling maximum plane subhaloes at 2.58 Gyr (left panel) and 7.73 Gyr (right panel) in the dark matter-only run. While there are wide, sheet-like infalling dark matter structures, the distinct and narrow filaments that are seen for h239 in Fig.~\ref{fig:h239g_filament} are not present. This configuration leads to a very transient, and non-coherent plane.

Further examples that strengthen the picture being painted above are found in the h285 dark matter-only and h285 baryonic runs. Recall from Fig.~\ref{fig:planes_hammer} that the maximum plane in the dark matter-only run is very transient, while the maximum plane in the baryonic run has a majority of its satellites coherently rotating. Fig.~\ref{fig:fig_h285_filament_00108} shows that, even though there are distinct filamentary accretion structures present, the maximum plane subhaloes in the dark matter-only h285 halo come in through all of them. On the other hand, we can see in Fig.~\ref{fig:fig_h285g_filament_00132} that even though the baryonic h285 has the exact same filamentary structures at a similar time, the majority of the coherent satellites of the maximum plane come in through two narrow filaments that are close to each other and another diametrically opposite filament, rather than through multiple as in the dark matter-only case.  
From these examples, it appears that maximum planes that accrete their satellites through multiple (more than two) sets of filaments do not lead to rotational coherence.  If, however, a maximum plane is built from satellites that are accreted primarily through two filaments (or two sets of filaments), this can lead to a more coherently rotating plane.  Previous work has noted that filamentary accretion can lead to planar satellite structures (see Introduction), but here we refine the results. Coherent rotation seems to more strongly arise if only two filaments contribute to the majority of planar satellites, as is the case in h239 and the baryonic h285 simulations.  While a maximum plane can still be defined in other cases, the plane does not contain a large number of coherently rotating subhaloes.  This is seen, for example, in the dark matter-only h258, where a large number of subhaloes fall in through many different filaments.  It is also seen in the dark matter-only h285, where amorphous accretion makes it very difficulty for the maximum plane to have any coherent rotational structure at $z=0$.

Finally, we note that the maximum plane is not aligned with the angular momentum vector of the dark matter halo or baryonic disc in any of our galaxies (this can be seen by the fact that the clustering of the satellite orbital angular momentum vectors in Fig.~\ref{fig:planes_hammer} do not cluster around the blue or red points).  As was noted in Fig.~\ref{fig:cosPDF}, the angular momentum of the baryonic disc in our simulations is always either aligned, or nearly anti-aligned in h285, with the dark matter halo angular momentum.  Despite that, the maximum plane is usually at an angle to them ($\psi$, listed in Table \ref{table:results}).

None of our planes are polar, like that of the Milky Way, but neither are they completely aligned with the disc.  As was mentioned before (see discussion of Fig.~\ref{fig:cosPDF}), generally the angular momentum vectors of \emph{all} the subhaloes (before selecting for a luminous population) do show preference to align (or anti-align) with the DM halo or baryonic disc.  In h258, the subhaloes were strongly aligned in the direction of the disc.  The satellite planes of the Milky Way and Andromeda are not aligned with their discs. We see here that, even if the majority of subhaloes tend to align with the disc (e.g., h258), the maximum plane does not. It is suggestive that something similar could be happening in real galaxies.

\section{Comparison of DM-only and DM+SPH simulations} \label{Comparison}

Being able to compare the same simulation with and without baryons lets us have a unique perspective on how baryons affect subhaloes and plane formation. Since our dark matter-only and SPH simulations start from the same initial conditions, it is natural to compare a subhalo from the dark matter-only simulation with its counterpart from the SPH simulation at $z=0$. To find counterparts, we track all the dark matter particles in a subhalo in the dark matter-only simulation and find which subhalo the majority of the same particles reside in the SPH simulation.

\begin{table}
\centering
\caption{Overlaps in subhaloes between DM and SPH simulations. Overlap number pairs are DM $\rightarrow$ SPH and SPH $\rightarrow$ DM correspondences respectively. See section \ref{Comparison} for details. $N_{destroyed}$ indicates the number of subhaloes that have been fully disrupted in the baryonic run that have surviving counterparts in the dark matter-only run.}

\begin{tabular}{ccccc}

\hline
 & h239 & h258 & h277 & h285 \\
\hline
$N_{sats}$ overlap & 18,20 & 11,16 & 13,14 & 5,9 \\
$N_{max}$ overlap & 4,4 & 1,3 & 2,3 & 2,2 \\
$N_{destroyed}$ & 2 & 1 & 4 & 10 \\

\hline
\label{table:overlap}

\end{tabular}
\end{table}

In our analysis, we check how many such correspondences occur in our total luminous subhalo selection, and in the maximum plane subhaloes. This is shown in Table \ref{table:overlap}. $N_{sats}$ overlap lists how many of the same subhaloes exist in the total luminous satellite population between DM and SPH simulations.  The two numbers are, first, the correspondence when starting from the dark matter-only run and tracing to the SPH run, and second, the correspondence when starting from the SPH run and tracing to dark matter-only run, respectively. $N_{max}$ overlap lists the same numbers, but for only the maximum plane subhaloes.

While our selection method picks out anywhere from 23 to 42 luminous satellites in the SPH run, and 30 in all dark matter-only runs, the number of overlapping haloes is anywhere from 5 to 20.  This difference is due to two facts.  First, the two methodologies used to identify luminous satellites (30 most massive at infall in the dark matter-only simulations, and all satellites with stellar mass above 2$\times$10$^4\ \mathrm{M_{\odot}}$ in the SPH runs) pick out slightly different subhaloes.  On top of that, the luminous population in the SPH simulation can be significantly different than the one in the dark matter-only simulation. In particular, a surviving subhalo in the dark matter-only run is more likely to be entirely disrupted in the SPH run \citep{Brooks2014}. The trend in the number of overlapping haloes in Table \ref{table:overlap} goes in this direction: we are less likely to find subhaloes in the SPH run that exist in the dark matter-only run than to find subhaloes in the dark matter-only run that exist in the SPH run.  This trend can be explained if the dark matter-only run contains subhaloes that have been fully destroyed in the SPH run. $N_{destroyed}$ in  Table \ref{table:overlap} indicates the number of subhaloes that have been fully disrupted in the baryonic run that have surviving counterparts in the dark matter-only run. 

In all four cases, the proportion of overlap in the luminous satellite populations (50-57 per cent) is greater than the proportion of overlap in the max plane population (7-30 per cent). This indicates that there are distinctly different maximum planes being picked out by the plane finding algorithm in the SPH simulations compared to the dark matter-only simulations. While the general statistics of the maximum plane stays relatively constant between the DM-only and DM+SPH simulations, the individual members of the planes are different. 
 
The presence of baryons changes the formation history, infall times and interactions with the parent haloes enough that by $z=0$, the dark matter-only and SPH versions of any given subhalo have different trajectories. This is in line with the findings of \citet{Schewtschenko2011}, who conclude that baryonic subhaloes have in general a later infall time than their dark matter-only counterparts, leading to different positions and velocities at $z=0$.

To test the robustness of this claim, we modified our selection criterion for luminous satellites and reran our plane detection algorithm.  In the SPH simulations we only chose satellites that corresponded to an existing one in the dark matter-only luminous population. Running the plane analysis algorithm on these also led to distinctly different maximum planes with minimal overlap between the dark matter-only and SPH planes (no more than 2-3 overlapping satellites in the maximum planes in all cases). Thus, the different planes identified between the two versions are not an effect of the method being used to select the luminous satellite population.  Instead, it is primarily the changes in formation and infall history introduced by the presence of baryons that leads to a different satellite configuration at $z=0$.  

The fact that the two versions of the simulations consistently lead to different maximum planes indicates that no study to date that has examined the formation of satellites planes using dark matter-only simulations has studied a realistic population of satellites.  The planes studied in dark matter-only simulations are not the same planes that result in the presence of baryons.  In order to study satellite planes with a realistic position, mass, and luminosity distribution, baryonic simulations are necessary.

Finally, we examined the trajectories of the satellites that are fully destroyed in the SPH run in order to test whether there is any geometrical dependence on the destruction, e.g., are satellites that have orbits aligned with the disc plane preferentially destroyed over other satellite orbits?  Indeed, we find that more than 75 per cent of all of the destroyed satellites have orbits that appear to be dragged toward the disc plane prior to their destruction.  Such disc-plane dragging was also demonstrated in \citet{Read2009}.  Another $\sim$25 per cent of the satellites instead seem to have radial paths that take them directly through the disc plane and they are quickly disrupted thereafter.  Intriguingly, all of the destroyed satellites have surviving counterparts in the dark matter-only run with positions at $z=0$ that are less than 100 kpc in height from the plane of the disc in the baryonic run, though they are not confined to this distance in other dimensions.  This seems to indicate that these counterparts have a preference for orbits that are on a path more closely aligned with the disc plane in the baryonic run.

\section{Conclusions} \label{Conclusion}

We have examined the impact of including baryonic physics in forming planes of satellites around Milky Way-mass galaxies, like the planes observed in both the Milky Way and Andromeda.  We have studied the satellite planes that form in both dark matter-only runs and baryonic runs of the same haloes, to explicitly assess the impact of baryons.  The presence of baryons has two effects.  First, it  changes the satellite composition of the resulting planes. The majority of satellites that contribute to a plane are different in the baryonic version of a simulation than in a dark matter-only version.  This is true even if the same satellites are used across both the baryonic and dark matter-only runs.  The resulting distribution of the satellites in the baryonic run is different enough that the maximum plane (the plane that maximizes the number of member satellites) is always different in the two versions.
However, the second effect of baryons is perhaps the most important in terms of understanding the significance of the planes observed around the Milky Way and Andromeda.  Inclusion of baryons makes the satellite distribution less radially concentrated (see Figure~\ref{fig:cumulative_subs}), leading to planes of satellites with higher significance compared to a random distribution.

To understand the effect of baryons on the planar significance, it is important to note that all of our planes pass through the centre of the halo, by definition.  Thus, as radial concentration decreases, it becomes much less likely to create a planar distribution when randomly populating the halo with satellites, even using the same radial distribution.   Because of this, the significance of planar structures increases compared to random when the satellites are less radially concentrated. When there are more centrally concentrated satellites, it becomes much easier to find as many (or more) satellites in the maximum plane in a random distribution. The presence of a disc in baryonic simulations tends to destroy satellites that pass near the centre of the galaxy (see Section~\ref{Comparison}, and Table \ref{table:overlap}, and this effect makes the planes in baryonic simulations more likely to be of higher significance.

Hence, it is critical to study simulations that include baryons if one wishes to understand the origin of highly significant planes.  Additionally, dark matter-only simulations lead to different plane satellite members, making it again critical to use baryonic simulations if one wishes to study the satellite members that contribute to significant planes.

Low radial concentration is part of the reason why the satellite plane in M31 has such high significance. M31 has a satellite distribution that is less concentrated near the galaxy centre.  In our simulations, we can find relatively low positional $p$-values due to low radial concentration.  However, to be as truly significant as the observed plane in Andromeda, we must also have a low kinematic $p$-value.  Although we have some simulations that produce low $p_{kin}$, we never find a case where both low $p_{kin}$ and low $p_{pos}$ occur together.  Hence, none of our simulated galaxies have planes as significant as M31 or the Milky Way.

In every simulation, we can easily identify a group of satellites that form a planar structure.  Notably, this plane is usually unique.  Our plane detection method consistently identifies a plane containing the maximum number of dwarfs.  Trying to define planes with different configurations leads to a plane with many fewer subhaloes. In this sense, it is easy to identify planes of satellites in simulations.  However, it is much more rare to be able to identify a plane with a large fraction of co-rotating satellites.  This is true despite the fact that we define co-rotation as a simple binary choice: the orbital angular momentum vectors of the satellites point to one side of the plane or the other.  When true co-rotation is examined (e.g., satellites whose orbital angular momentum vectors cluster together near the poles in Fig.~\ref{fig:planes_hammer}), the number of coherently rotating satellites is even smaller.  In reality, all of the defined planes contain transient satellites that appear to be spatially coincident with the plane, but that have an orbit that will eventually move them out of the plane.  Even our most coherently rotating planes contain some transients, and in some cases the entire plane can be made of transients.

We also investigate the role of large-scale filamentary structure in forming satellite planes. For those planes that do contain a significant fraction of co-rotating satellites, the coherently rotating satellites seem to be accreted primarily through no more than two sets of filaments.  When more than two sets of filaments contribute, or if there are no well-defined filaments in general, then coherent rotation of the plane does not exist.

Overall, a simulated galaxy must be able to produce a satellite plane that is both positionally significant (with low radial concentration) and kinematically significant (with more than $\sim$80 per cent co-rotating) in order to match the highly significant satellites planes observed around the Milky Way and Andromeda.  Such a simulation has yet to be published.

\section*{Acknowledgements}
AMB and SHA acknowledge support from NSF award AST-1411399.  Resources supporting this work were provided by the NASA High-End Computing (HEC) Program through the NASA Advanced Supercomputing (NAS) Division at Ames Research Center.  We thank Tom Quinn and James Wadsley for use of the proprietary code {\sc Gasoline}, and Fabio Governato for help in producing the initial conditions and running the simulations. The pynbody package \citep{pynbody} and SciPy package \citep{scipy} were used in portions of this analysis.

\appendix 
\section{Details of Satellite Selection Criteria} \label{Appendix}

In addition to the methods mentioned in Section \ref{IdentifyingSatellites}, we tested various selection criteria for choosing luminous satellites in both our DM-only and DM+SPH simulations. 

For the DM-only simulations, we alternatively tried selecting the 30 most massive subhaloes at $z=0$ instead of the 30 most massive at infall. 
Roughly 80 per cent of the most massive subhaloes at $z=0$ are also in the sample selected by 30 most massive at infall.  
We also investigated the effect that using as few as 25 subhaloes and as many as 40 subhaloes had on our results. 
While changing the total number selected does change $N_{sats}$ and $N_{max}$ proportionally, in all cases the alternative selections have no significant effect on our $p$-values nor the conclusions drawn about filamentary accretion. 

Alternatively, we tested selecting subhaloes with a $v_{max}$ (maximum value of the rotation curves) at $z=0$ $>15.0\ \mathrm{km\ s^{-1}}$ in the dark matter-only runs, and identified the surviving counterparts in the baryonic run.  This selection is identical to that used in \citet{Zolotov2012} and \citet{Brooks2014}, which yielded a satellite sample that produced realistic luminosity functions and velocity dispersions.  However, this method was designed originally to capture primarily classically bright dwarf spheroidals ($M_V$ brighter than $-8$ mag).  It led to too few luminous dwarfs compared to the number observed around the Milky Way and M31.  Again, we note that even in the case of requiring the dark matter-only and baryonic satellites to be counterparts of each other, the resulting planes are different due to the different infall times and orbital evolution in the two runs. 

In the baryonic runs, we are able to select luminous satellites directly.  However, as the number of star particles decreases, the star formation history in any individual galaxy is less likely to be converged.  Subhaloes with more than 5 star particles are more likely to have converged star formation histories.  However, selecting satellites with more than 5 star particles decreased the number of luminous satellites by up to 10 in most of the galaxies, making the numbers far fewer than observed in the Milky Way or M31.  We settled on a minimum of 4 star particles (corresponding to a lower stellar mass cutoff of $M_{star}>2\times10^4\ \mathrm{M_{\odot}}$) as a compromise that yielded a reasonable number of luminous satellites. Decreasing to 3 or more star particles resulted in adding anywhere between 8-20 more subhaloes, yielding more satellites than we needed and resulting in a questionable convergence of the star formation history in the least luminous satellites.

%\pagebreak
%\newpage
\bibliographystyle{mnras}
\bibliography{references}

%\Comment{Not all text citations linked in bibliography}

\bsp	% typesetting comment
\label{lastpage}
\end{document}